\documentclass[aps,prl,showpacs,showkeys,amsmath,amssymb,
twocolumn,
floatfix,
superscriptaddress
]{revtex4-1}
\usepackage{graphicx}
\usepackage{times}
\usepackage{verbatim}
\usepackage{color}
\usepackage{bm}
\usepackage{natbib}
\usepackage[normalem]{ulem}
\definecolor{gray}{rgb}{0.05,0.05,0.55}
\usepackage[colorlinks,linkcolor=gray,anchorcolor=gray,citecolor=gray, urlcolor=gray,filecolor=gray]{hyperref}
\usepackage{array}

\graphicspath{{Figures/}}
\usepackage{graphicx}% Include figure files

\begin{document}

%\preprint{APS/123-QED}

\title{Joint Determination of Reactor Antineutrino Spectra from $^{235}$U and $^{239}$Pu Fission by Daya Bay and PROSPECT}

%%%% Generator info for file
%%%% Thu Apr  8 09:27:29 2021
%%%% /Users/djaffe/Documents/Reactor/work/trunk/NuWa-trunk/dybgaudi/Documentation/AuthorList_SAVE/2021_PROSPECT_DYB_Spectrum
%%%% merge.py 
%
\newcommand{\NUU}{\affiliation{National~United~University, Miao-Li}}
\newcommand{\NIST}{\affiliation{National Institute of Standards and Technology, Gaithersburg, MD, USA}}
\newcommand{\UCB}{\affiliation{Department of Physics, University~of~California, Berkeley, California  94720, USA}}
\newcommand{\Hawaii}{\affiliation{Department of Physics \& Astronomy, University of Hawaii, Honolulu, HI, USA}}
\newcommand{\IIT}{\affiliation{Department of Physics, Illinois Institute of Technology, Chicago, IL, USA}}
\newcommand{\SJTU}{\affiliation{Department of Physics and Astronomy, Shanghai Jiao Tong University, Shanghai Laboratory for Particle Physics and Cosmology, Shanghai}}
\newcommand{\HKU}{\affiliation{Department of Physics, The~University~of~Hong~Kong, Pokfulam, Hong~Kong}}
\newcommand{\UCI}{\affiliation{Department of Physics and Astronomy, University of California, Irvine, California 92697, USA}} %20190209 for Pedro's new digs
\newcommand{\TsingHua}{\affiliation{Department~of~Engineering~Physics, Tsinghua~University, Beijing}}
\newcommand{\VirginiaTech}{\affiliation{Center for Neutrino Physics, Virginia~Tech, Blacksburg, Virginia  24061, USA}}
\newcommand{\ORNL}{\affiliation{Physics Division, Oak Ridge National Laboratory, Oak Ridge, TN, USA}}
\newcommand{\NJU}{\affiliation{Nanjing~University, Nanjing}}
\newcommand{\ECUST}{\affiliation{Institute of Modern Physics, East China University of Science and Technology, Shanghai}}
\newcommand{\CUHK}{\affiliation{Chinese~University~of~Hong~Kong, Hong~Kong}}
\newcommand{\UIUC}{\affiliation{Department of Physics, University~of~Illinois~at~Urbana-Champaign, Urbana, Illinois 61801, USA}}
\newcommand{\Drexel}{\affiliation{Department of Physics, Drexel University, Philadelphia, PA, USA}}
\newcommand{\HFIR}{\affiliation{High Flux Isotope Reactor, Oak Ridge National Laboratory, Oak Ridge, TN, USA}}
\newcommand{\BNL}{\affiliation{Brookhaven National Laboratory, Upton, NY, USA}}
\newcommand{\NUDT}{\affiliation{College of Electronic Science and Engineering, National University of Defense Technology, Changsha}} % added 20140111
\newcommand{\CIAE}{\affiliation{China~Institute~of~Atomic~Energy, Beijing}}
\newcommand{\TENN}{\affiliation{Department of Physics and Astronomy, University of Tennessee, Knoxville, TN, USA}}
\newcommand{\GXU}{\affiliation{Guangxi University, No.100 Daxue East Road, Nanning}} % added 20201208 ,530004, China
\newcommand{\BNU}{\affiliation{Beijing~Normal~University, Beijing}}
\newcommand{\Iowa}{\affiliation{Iowa~State~University, Ames, Iowa  50011, USA}}
\newcommand{\ZSU}{\affiliation{Sun Yat-Sen (Zhongshan) University, Guangzhou}}
\newcommand{\NanKai}{\affiliation{School of Physics, Nankai~University, Tianjin}}
\newcommand{\TempleUniversity}{\affiliation{Department of Physics, Temple University, Philadelphia, PA, USA}}
\newcommand{\GATECH}{\affiliation{George W.\,Woodruff School of Mechanical Engineering, Georgia Institute of Technology, Atlanta, GA USA}}
\newcommand{\DGUT}{\affiliation{Dongguan~University~of~Technology, Dongguan}}
\newcommand{\UC}{\affiliation{Department of Physics, University~of~Cincinnati, Cincinnati, Ohio 45221, USA}}
\newcommand{\BCC}{\altaffiliation{Now at Department of Chemistry and Chemical Technology, Bronx Community College, Bronx, New York  10453, USA}} % add 20160311 for Sunej's 'now at'
\newcommand{\NCTU}{\affiliation{Institute~of~Physics, National~Chiao-Tung~University, Hsinchu}}
\newcommand{\XJTU}{\affiliation{Department of Nuclear Science and Technology, School of Energy and Power Engineering, Xi'an Jiaotong University, Xi'an}}%%updated 20161215 {Xi'an Jiaotong University, Xi'an}}
\newcommand{\LLNL}{\affiliation{Nuclear and Chemical Sciences Division, Lawrence Livermore National Laboratory, Livermore, CA, USA}}
\newcommand{\Waterloo}{\affiliation{Institute for Quantum Computing and Department of Physics and Astronomy, University of Waterloo, Waterloo, ON, Canada}}
\newcommand{\Berkeley}{\affiliation{Lawrence~Berkeley~National~Laboratory, Berkeley, California 94720, USA}}
\newcommand{\Charles}{\affiliation{Charles~University, Faculty~of~Mathematics~and~Physics, Prague, Czech Republic}} %updated20140819
\newcommand{\CGNPG}{\affiliation{China General Nuclear Power Group, Shenzhen}}% update 20170729 add city. updated 20140724 China~Guangdong~Nuclear~Power~Group, Shenzhen}}
\newcommand{\Yale}{\affiliation{Wright Laboratory, Department of Physics, Yale University, New Haven, CT, USA}}
\newcommand{\Wisconsin}{\affiliation{Department of Physics, University of Wisconsin, Madison, Madison, WI, USA}}
\newcommand{\Dubna}{\affiliation{Joint~Institute~for~Nuclear~Research, Dubna, Moscow~Region, Russia}}
\newcommand{\LeMoyne}{\affiliation{Department of Physics, Le Moyne College, Syracuse, NY, USA}}
\newcommand{\NTU}{\affiliation{Department of Physics, National~Taiwan~University, Taipei}}
\newcommand{\Princeton}{\affiliation{Joseph Henry Laboratories, Princeton~University, Princeton, New~Jersey 08544, USA}}
\newcommand{\SDU}{\affiliation{Shandong~University, Jinan}}
\newcommand{\CNIHEP}{\affiliation{Institute~of~High~Energy~Physics, Beijing}}
\newcommand{\CQU}{\affiliation{Chongqing University, Chongqing}} % add 20150417
\newcommand{\Siena}{\affiliation{Siena~College, Loudonville, New York  12211, USA}}
\newcommand{\NCEPU}{\affiliation{North~China~Electric~Power~University, Beijing}}
\newcommand{\Caltech}{\affiliation{California~Institute~of~Technology, Pasadena, California 91125, USA}}
\newcommand{\WandM}{\affiliation{College~of~William~and~Mary, Williamsburg, Virginia  23187, USA}}
\newcommand{\SZU}{\affiliation{Shenzhen~University, Shenzhen}}
\newcommand{\USTC}{\affiliation{University~of~Science~and~Technology~of~China, Hefei}}
\author{F.~P.~An\ensuremath{^{\delta}}}
\ECUST 
\author{M.~Andriamirado\ensuremath{^{\pi}}}
\IIT 
\author{A.~B.~Balantekin\ensuremath{^{\delta}}\ensuremath{^{\pi}}}
\Wisconsin 
\author{H.~R.~Band\ensuremath{^{\pi}}}
\Yale 
\author{C.~D.~Bass\ensuremath{^{\pi}}}
\LeMoyne 
\author{D.~E.~Bergeron\ensuremath{^{\pi}}}
\NIST 
\author{D.~Berish\ensuremath{^{\pi}}}
\TempleUniversity 
\author{M.~Bishai\ensuremath{^{\delta}}}
\BNL 
\author{S.~Blyth\ensuremath{^{\delta}}}
\NTU 
\author{N.~S.~Bowden\ensuremath{^{\pi}}}
\LLNL 
% ...
\collaboration{\ensuremath{^{\delta}}Daya Bay Collaboration}\noaffiliation
\collaboration{\ensuremath{^{\pi}}PROSPECT Collaboration}\noaffiliation

%\author{Daya Bay Collaboration \& PROSPECT Collaboration}

\date{\today}% It is always \today, today,

\begin{abstract}
A joint determination of the reactor antineutrino spectra resulting from the fission of $^{235}$U and $^{239}$Pu has been carried out by the Daya Bay and PROSPECT collaborations.
This Letter reports the level of consistency of $^{235}$U spectrum measurements from the two experiments and presents new results from a joint analysis of both data sets.
The measurements are found to be consistent.
The combined analysis reduces the degeneracy between the dominant $^{235}$U and $^{239}$Pu isotopes and improves the uncertainty of the $^{235}$U spectral shape to about 3\%. 
The ${}^{235}$U and $^{239}$Pu antineutrino energy spectra are unfolded from the jointly deconvolved reactor spectra using the Wiener-SVD unfolding method, providing a data-based reference for other reactor antineutrino experiments and other applications. 
This is the first measurement of the $^{235}$U and $^{239}$Pu spectra based on the combination of experiments at low- and highly enriched uranium reactors.
\end{abstract}

\maketitle

During the operation of low-enriched uranium (LEU) commercial reactors, electron antineutrinos ($\bar{\nu}_e$) are emitted through the beta decays of fragments generated by the fissions of ${}^{235}$U, ${}^{238}$U, ${}^{239}$Pu, and ${}^{241}$Pu.  
Predictions of the $\bar{\nu}_e$~energy spectra produced by these fission isotopes have been generated via conversion of aggregate fission beta spectrum measurements~\cite{Schreckenbach:1985ep,VonFeilitzsch:1982jw,Hahn:1989zr,Mueller:2011nm,Huber:2011wv,bib:munich} or via summation of $\bar{\nu}_e$ contributions from all individual beta decay branches using standard nuclear databases~\cite{Mueller:2011nm, Fallot:2012jv, Dwyer:2014eka,Estienne:2019ujo}.  
Many significant neutrino physics measurements, such as the discovery of the neutrino~\cite{cowan1956}, the determination of neutrino mass differences and flavor mixing amplitudes~\cite{KamLAND_rate, KamLAND_shape, bib:dc, DoubleChooz:2019qbj, bib:reno, bib:reno_shape, Bak:2018ydk, bib:prl_rate, bib:prl_shape, Adey:2018zwh}, and searches for active-to-sterile neutrino oscillations~\cite{bib:Bugey3_osc, prospect_osc, stereo_2018, danss_osc,Adamson:2020jvo,Choi:2020ttv,Serebrov:2018vdw}, have used relatively little knowledge of these isotopic reactor $\bar{\nu}_e$ spectra.  
However, future reactor-based efforts probing important neutrino properties, such as the mass ordering~\cite{juno1, juno2} and coherent neutrino-nucleus scattering cross-sections~\cite{CONNIE, CONUS, RICOCHET, MINER, NUCLEUS}, may rely on a detailed and accurate understanding of $\bar{\nu}_e$ energy spectra and fluxes.  
Moreover, a variety of $\bar{\nu}_e$-based safeguard efforts~\cite{Bernstein:2019hix,PhysRevApplied.9.014003,PhysRevApplied.8.034005,Stewart:2019rtd} and nuclear data validations~\cite{bib:IAEA} are reliant on proper understanding of $\bar{\nu}_e$ emissions from different reactor types and fuel compositions.  

Reactor $\bar{\nu}_e$ can be measured in organic scintillator via the inverse beta decay (IBD) reaction: $\bar{\nu}_e + p \rightarrow e^+ + n$.  
Energy deposited by an IBD positron and its subsequent annihilation gammas form a prompt scintillation signal that is used to determine the kinetic energy of the interacting $\bar{\nu}_e$. 
In recent years, several reactor $\bar{\nu}_e$ measurements have cast doubt on the accuracy of existing conversion and summation predictions.  
Specifically, prediction-data tensions have been reported for both LEU reactor $\bar{\nu}_e$ fluxes~\cite{Mention:2011rk,An:2017osx,Atif:2020eyo,An:2016srz,DoubleChooz:2019qbj} and energy spectra~\cite{An:2015nua,An:2016srz,RENO:2015ksa,Abe:2014bwa,Ko:2016owz,Zacek:2018bij}. 
Conclusions of prediction-data disagreement in measurements from LEU reactors are similar whether results are reported in terms of $\bar{\nu}_e$ energy \cite{An:2016srz, Atif:2020eyo,DayaBay:2021dqj} or in terms of reconstructed energy from $\bar{\nu}_e$ signals~\cite{Adey:2019ywk, Atif:2020eyo}.

Additional reactor $\bar{\nu}_e$ measurements have been performed to evaluate the role played by each individual fission isotope in generating these observed discrepancies.
By exploiting variations in its reactors' fuel content and using conservative assumptions about $\bar{\nu}_e$ contributions from sub-dominant fission isotopes ${}^{238}$U~and ${}^{241}$Pu, the Daya Bay experiment has extracted prompt energy spectra from LEU reactors for the dominant fission isotopes ${}^{235}$U~and ${}^{239}$Pu~\cite{Adey:2019ywk}.  
In the 4-6~MeV prompt energy region, the greatest relative contribution to the prediction-data spectral shape disagreement, Daya Bay measures a 7\% (9\%) excess of events in ${}^{235}$U~(${}^{239}$Pu) relative to Huber-Mueller conversion predictions~\cite{Huber:2011wv,Mueller:2011nm} with an IBD cross-section applied~\cite{Vogel:1999zy}. Here the ${}^{238}$U component makes up about 8\% of the total fission for Daya Bay reactors and its prediction is based on the summation model from Mueller~\cite{Mueller:2011nm}.
To facilitate a comparison of spectral shapes, the predictions are scaled to the same integrated rate as the measurements.
Meanwhile, the PROSPECT experiment has performed a pure $^{235}$U prompt energy spectrum measurement using $\bar{\nu}_e$ fluxes from a highly enriched uranium (HEU) compact research reactor core \cite{Ashenfelter:2018jrx,Andriamirado:2020erz}. 
PROSPECT’s spectrum measurement also shows a prediction-data disagreement consistent with those observed by the LEU-based experiments. 
A recent measurement of spectral shape at an HEU core by the STEREO experiment indicates similar conclusions~\cite{AlmazanMolina:2020jlh}. 

This Letter evaluates the consistency of measured prompt energy spectra attributed to $\bar{\nu}_e$ from $^{235}$U fission with the Daya Bay and PROSPECT experiments. 
With consistency of derived spectra assured, a joint analysis of both experiments' data improves the precision of the derived $^{235}$U~spectrum and reduces the degeneracy between derived ${}^{235}$U~and ${}^{239}$Pu~spectra below that of a standalone analysis. 
The $\bar{\nu}_e$ energy spectra of ${}^{235}$U and ${}^{239}$Pu are then unfolded with the Wiener-SVD method~\cite{Tang:2017rob}, providing more precise data-based predictions than previously available for other reactor $\bar{\nu}_e$ experiments. 

The Daya Bay experiment measures $\bar{\nu}_e$ from the Daya Bay nuclear power complex, which hosts six 2.9 GW thermal power LEU commercial pressurized-water reactors~\cite{Guo:2007ug}. 
Eight identically designed antineutrino detectors (ADs)~\cite{DayaBay:2012aa,An:2015qga,Band:2013osa} are deployed in two near halls (two ADs each) and one far hall (four ADs).  
Each AD consists of a stainless steel tank with two nested cylindrical acrylic vessels~\cite{Band:2012dh}. 
The inner vessel contains 20~tons of 0.1\% Gd-loaded liquid scintillator (GdLS)~\cite{Beriguete:2014gua}, which serves as the active $\bar{\nu}_e$ detector volume for IBD reactions.  
The outer vessel holds a 42~cm thick layer of pure liquid scintillator (LS) region to improve detection of gamma rays escaping from the GdLS region.  
The IBD prompt signal is followed by an energy deposition from Gd-capture of the IBD neutron approximately 30 $\mu$s later on average. 
This Letter uses 3.5 million IBD events observed by four near hall ADs in combination with reactor fission fraction evolution to extract $^{235}$U and $^{239}$Pu spectra ~\cite{Adey:2019ywk}.
Further details can be found in Refs.~\cite{An:2015qga,An:2016ses,Adey:2018qct,Adey:2018zwh}.  

The PROSPECT experiment measures $\bar{\nu}_e$ from the High Flux Isotope Reactor (HFIR, an HEU reactor) with 85~MW thermal power at Oak Ridge National Laboratory~\cite{Ashenfelter:2018zdm,Ashenfelter:2015uxt}.
A 4-ton $^6$Li-loaded liquid scintillator (LiLS) detector is deployed 7.9~m away from the reactor to detect $\bar{\nu}_e$ IBD interactions. 
To measure $\bar{\nu}_e$ with different baselines, the LiLS target is divided into an 11$\times$14 array of long, optically isolated, rectangular segments~\cite{Ashenfelter:2019lbf}.
IBD prompt signals are identified by their time correlation with the signal of an IBD neutron capture on $^6$Li, with over 99\% of the $\bar{\nu}_e$ produced in the fuel from HFIR due to $^{235}$U fission. 
This analysis uses 50,000 IBD events observed by PROSPECT~\cite{Andriamirado:2020erz} which are measured without absolute rate normalization.  
Further information can be found in Refs.~\cite{Ashenfelter:2018zdm,Ashenfelter:2018jrx,Andriamirado:2020erz}.  

\begin{figure}
    \centering
    \includegraphics[width=\columnwidth]{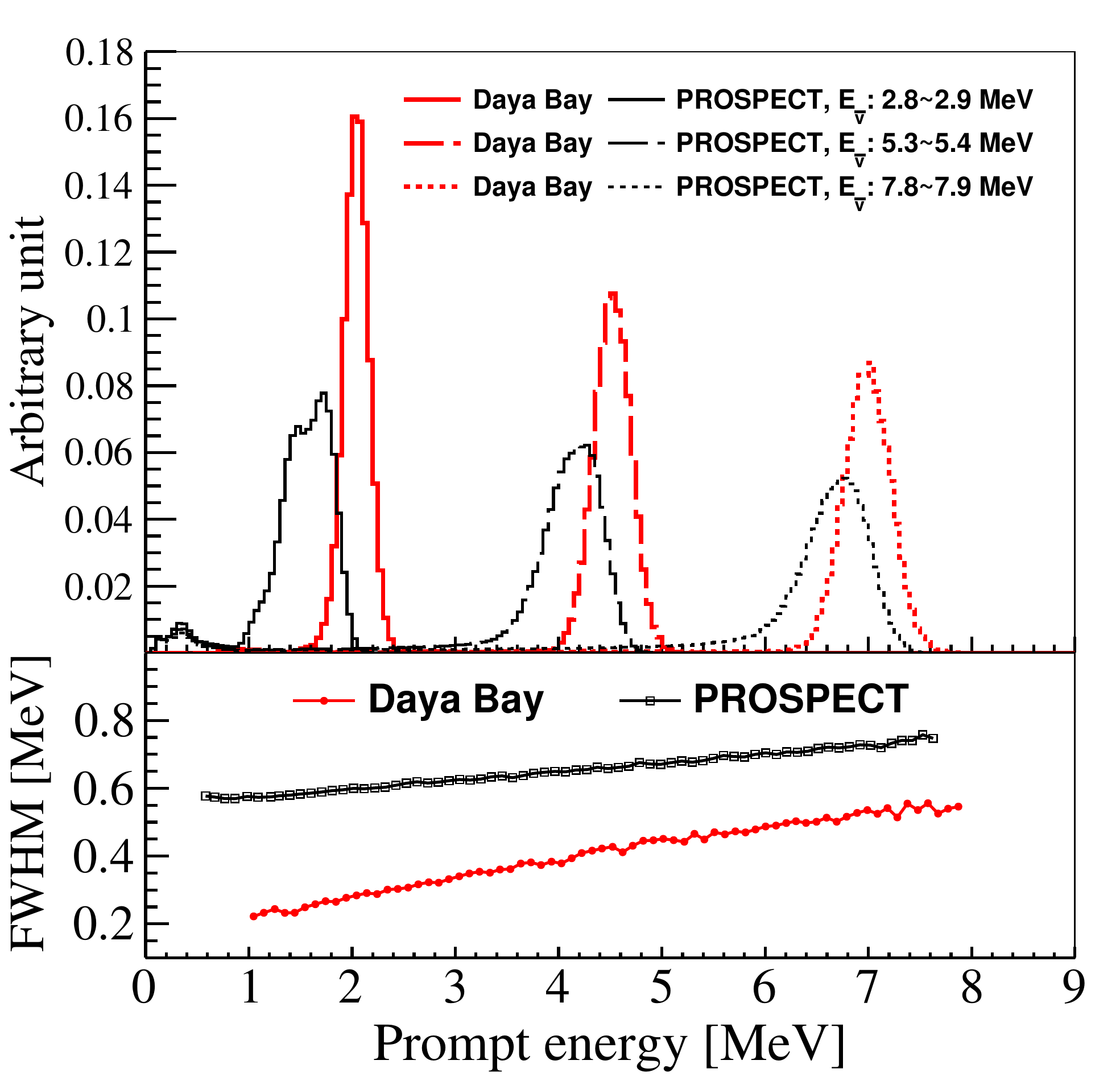}
    \caption{(Top) Reconstructed prompt energy distributions based on simulated $\bar{\nu}_e$ signals with specific $\bar{\nu}_e$ energy ranges (uniform distribution). 
    The areas of the distributions are normalized to 1. 
    The shift in peak location between the two experiments is driven primarily by the handling of scintillator non-linearity in the energy response. 
    These effects for PROSPECT are incorporated into the response function while this effect is taken into account by Daya Bay’s calibration methodology.
    (Bottom) FWHM of reconstructed prompt energy distributions versus prompt energy at the peak of those distributions. The difference in FWHM is primarily due to various effects from inactive volume in the detector response functions.  }
    \label{fig:ResponseDifferences}
\end{figure}

The different detector designs and energy reconstruction approaches between the two experiments resulted in distinct detector responses.
Examples of the reconstructed prompt energy distributions based on artificial $\bar{\nu}_e$ signals of distinct energies are shown in Fig.~\ref{fig:ResponseDifferences}. 
For Daya Bay, reconstructed prompt energies of IBD events are corrected for well-calibrated energy non-linearity and spatial non-uniformity effects, while in the compact and segmented PROSPECT detector the combined effects of energy non-linearity and leakage of prompt energy are included in a detector energy response function.  
The full width at half maximum (FWHM) of the reconstructed prompt energy distributions is shown in Fig.~\ref{fig:ResponseDifferences}. 
Even though the photo-statistics energy resolution of PROSPECT (4.8\% at 1~MeV) is better than Daya Bay's (7.8\% at 1~MeV), the total smearing of the full PROSPECT detector response is larger due to greater prompt energy leakage into inactive volumes within its detector.  
% 165 p.e./MeV for Daya Bay
 
\begin{figure}
    \centering
    \includegraphics[width=\linewidth]{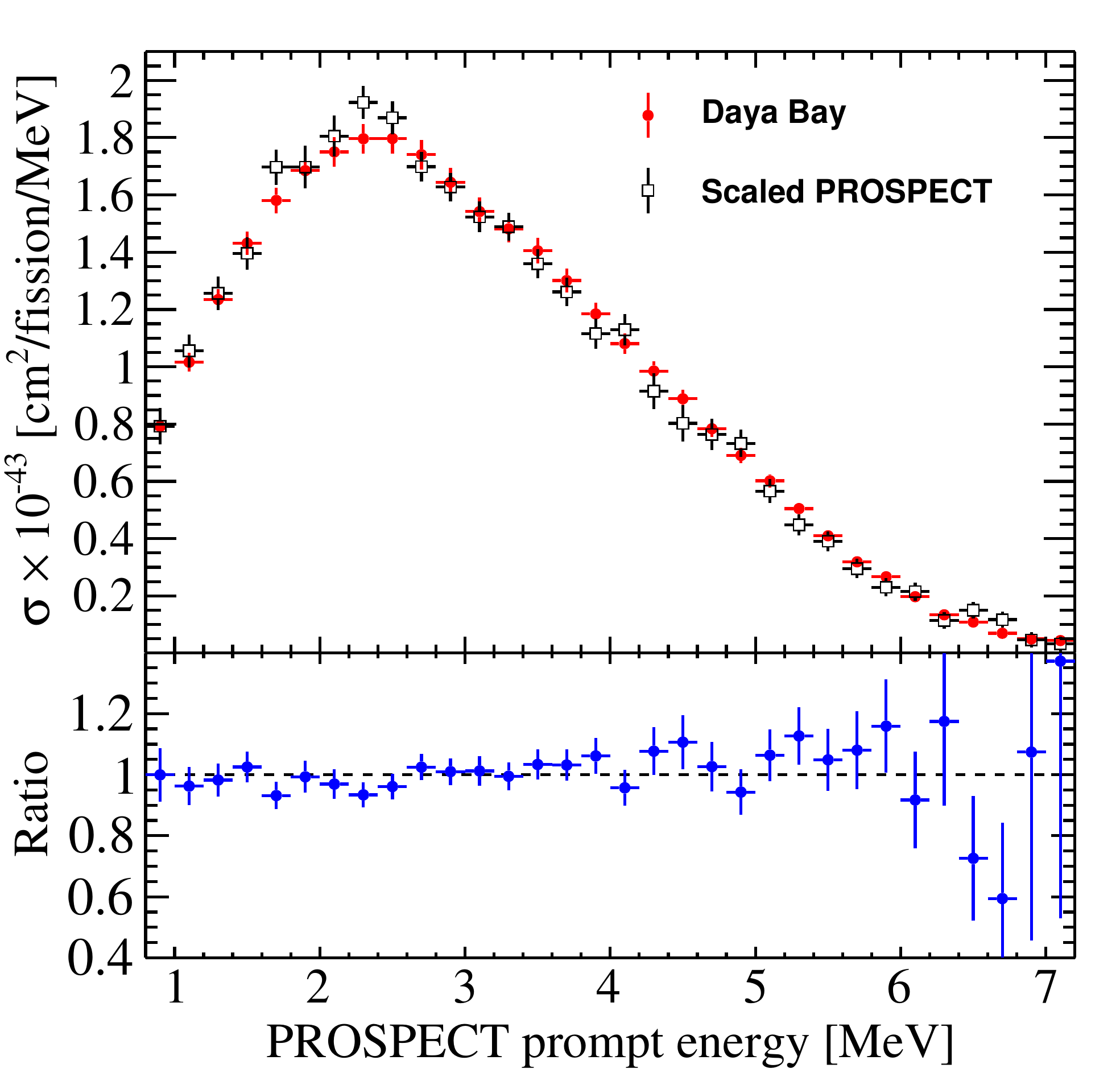}
    \caption{ Comparison of the measurements of the $^{235}$U prompt energy spectrum from Daya Bay and PROSPECT (top) and the ratio of the spectrum from Daya Bay over the one from PROSPECT (bottom). Here the measurement from Daya Bay has been transformed to the reconstructed energy scale of PROSPECT based on a dedicated response matrix $\boldsymbol{R}^{\rm map}$ and the y-axis has been scaled to match the absolute rate from the Daya Bay measurement. Error bars contain both statistical and systematic contributions. The measurements from Daya Bay and PROSPECT are consistent with each other.
    }
    \label{fig:PromptComparison}
\end{figure}

To assess the consistency of the ${}^{235}$U spectrum measurements by Daya Bay and PROSPECT, the spectra are converted into a common energy scale. 
The measured prompt energy spectrum $\boldsymbol{S}^{\rm e}_{\rm p}$ (where e = DYB (Daya Bay) or PRO (PROSPECT)) is the convolution of the detector response with the original $\bar{\nu}_e$ energy spectrum of ${}^{235}$U ($\boldsymbol{S}_{\bar{\nu}_e}$): 
\begin{equation}
\boldsymbol{S}^{\rm e}_{\rm p}= \boldsymbol{R}^{\rm e}\boldsymbol{S}_{\bar{\nu}_e}, 
\label{Eq-ResponseM}
\end{equation}
where $\boldsymbol{R}^{\rm e}$ is the $\bar{\nu}_e$ energy response function with e = DYB or PRO. 
To compare $\boldsymbol{S}^{\rm DYB}_{\rm p}$ and $\boldsymbol{S}^{\rm PRO}_{\rm p}$, a mapping matrix $\boldsymbol{R}^{\rm map}$ is constructed to transform the measured prompt energy spectrum of $^{235}$U at Daya Bay $\boldsymbol{S}^{\rm DYB}_{\rm p}$ to the corresponding spectrum $\boldsymbol{S}^{\rm DYB}_{\rm map}$ with the PROSPECT detector response: 
\begin{equation} 
\label{formula:mapmatrix}
\boldsymbol{S}^{\rm DYB}_{\rm map}=\boldsymbol{R}^{\rm map}\boldsymbol{S}^{\rm DYB}_{\rm p}=\boldsymbol{R}^{\rm PRO}(\boldsymbol{R}^{\rm DYB})^{-1}\boldsymbol{S}^{\rm DYB}_{\rm p}. 
\end{equation} 
The transformation to the energy space with poorer overall energy smearing avoids amplifying statistical fluctuations introduced in the unfolding procedure~\cite{Hocker:1995kb}. 
%This method is checked via dedicated toy Monte Carlo tests.
%In these tests, 10$^{4}$ Daya Bay and PROSPECT toy spectra generated with identical models were compared in the way shown below and found to have negligible bias. 
Although the Huber-Mueller model is used in the generation of $\boldsymbol{R}^{\rm e}$, the choice of model is found to have negligible impact on the construction of the mapping matrix and $\boldsymbol{S}^{\rm DYB}_{\rm map}$.
The comparison between $\boldsymbol{S}^{\rm DYB}_{\rm map}$ and the PROSPECT measurement ($\boldsymbol{S}^{\rm PRO}_{\rm p}$) is shown in Fig.~\ref{fig:PromptComparison}, where the PROSPECT measurement is normalized to the flux measured from Daya Bay. 
The error bars in the figure are the square root of the diagonal elements of the full covariance matrices, containing both statistical and systematic contributions. 
The lower panel incorporates uncertainties from both experiments in its error bars, and the measurements are consistent across the full energy range.

To further evaluate the consistency between Daya Bay and PROSPECT quantitatively, a $\chi^2$ function is constructed with a rate free parameter ($\eta^{\rm rate}$) instead of the enforced normalization:

\begin{align} 
\label{Eq:fitter}
\chi^2 =&\chi^2_{\rm DYB}+\chi^{2}_{\rm PRO} \nonumber\\
=&(\boldsymbol{S}^{\rm fit}-\boldsymbol{S}^{\rm DYB}_{\rm p})^{T} (\textbf{Cov}^{\rm DYB})^{-1} (\boldsymbol{S}^{\rm fit}-\boldsymbol{S}^{\rm DYB}_{\rm p})\nonumber\\
&+ (\boldsymbol{R}^{\rm map}\boldsymbol{S}^{\rm fit}\eta^{\rm rate}-\boldsymbol{S}^{\rm PRO}_{\rm p})^{T}(\textbf{Cov}^{\rm PRO})^{-1} \nonumber\\
&(\boldsymbol{R}^{\rm map}\boldsymbol{S}^{\rm fit}\eta^{\rm rate}-\boldsymbol{S}^{\rm PRO}_{\rm p}). 
\end{align} 

Here, $\boldsymbol{S}^{\rm fit}_i=\boldsymbol{H}^{\rm 235}_i\times\boldsymbol{\eta}_i$, $\boldsymbol{\eta}$ is a vector of free parameters to fit each prompt energy bin (with index $i$) of a common initial prediction of $^{235}$U spectrum $\boldsymbol{H}^{\rm 235}$ for both experiments. 
$\textbf{Cov}^{\rm DYB}$ and $\textbf{Cov}^{\rm PRO}$ are the covariance matrices of the measurements for Daya Bay \cite{Adey:2019ywk} and PROSPECT \cite{Andriamirado:2020erz}, respectively. 
Without the inclusion of PROSPECT data, the minimum $\chi^2$ would be 0. 
Based on the measurements from both experiments, the minimum $\chi^2$ is 25.44 with 31 degrees of freedom, corresponding to a $p$-value of 0.75. 
This result is further validated with a frequentist approach using the minimized $\chi^2$ values based on Eq.~(\ref{Eq:fitter}) from $10^{4}$ toy Monte Carlo tests, and the distribution of the $\chi^2$ values matches the $\chi^2$ distribution with 31 degrees of freedom as expected. 
Overall, the measured Daya Bay and PROSPECT $^{235}$U spectra are consistent with one another. 
\iffalse
To more precisely quantify the level of consistency between $^{235}$U spectra, a frequentist approach is employed. 
Using the covariance matrices from both experiments, $10^{4}$ toy Monte Carlo tests are generated. 
A probability distribution function (PDF) for $\chi^2$ is formed from the minimized $\chi^2$ values for each toy Monte Carlo test.
Using this PDF, a $p$-value of 0.782 is found for the consistency of the two spectra. 
\fi
Next, the significance of local deviations between the two spectra is evaluated by introducing an additional free parameter for each bin in 1~MeV wide sliding energy windows of one experiment such that the original test is a nested hypothesis of the new fit.
The significance of the difference in minimum $\chi^2$ before and after introducing these free parameters gives $p$-values all greater than 0.25, corresponding to local deviations less than 1.1$\sigma$ for all energy windows.

\begin{figure}
    \centering
    \includegraphics[width=\linewidth]{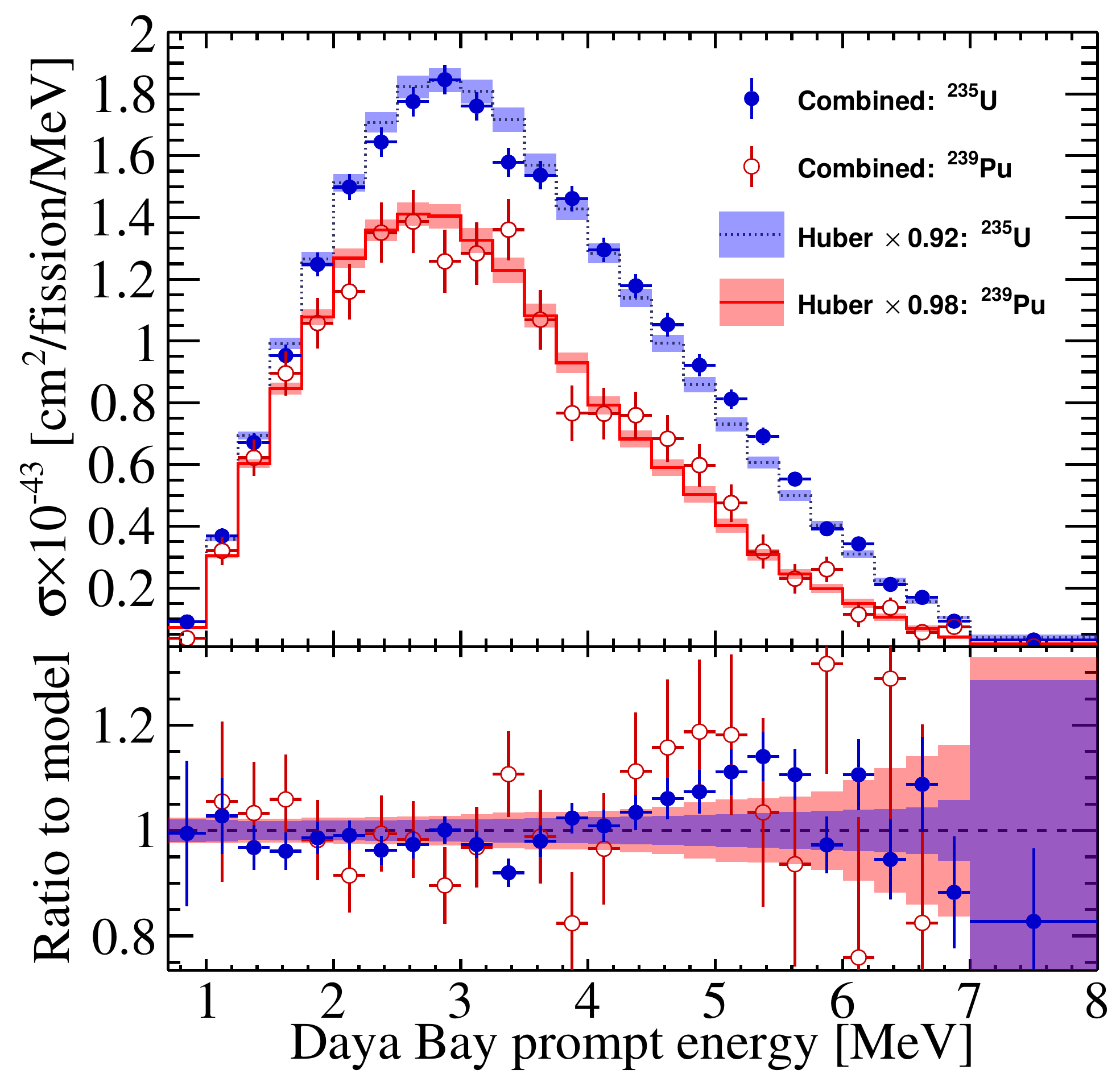}
    \caption{(Top) The extracted ${}^{235}$U and ${}^{239}$Pu spectra in Daya Bay's prompt energy from the combined analysis of the Daya Bay and PROSPECT data. The corresponding scaled Huber model predictions are overlaid. The error bars in the data points are the square root of the diagonal elements of the covariance matrix for the spectral shape, with no absolute rate uncertainty. (Bottom) The ratio of the combined analysis results to the shape predictions from the scaled Huber-Mueller model. }
    \label{fig:JointAnalysisPrompt}
\end{figure}
With no evidence of inconsistency between the two experiments, the PROSPECT measurement is incorporated in a joint fit $\chi^2 = \chi^{\prime 2}_{\textrm{DYB}} + \chi^2_{\textrm{PRO}}$ to improve the extraction of the ${}^{235}$U and ${}^{239}$Pu spectra in Daya Bay using the evolution of the prompt energy spectrum as a function of fission fractions~\cite{Adey:2019ywk}. 
To avoid additional uncertainties from the unfolding method mentioned above, the fit is done on the prompt energy spectra rather than the $\bar{\nu}_e$ energy spectra. 
In the joint fit, $\chi^{\prime 2}_{\textrm{DYB}}$ is the same as described in Ref.~\cite{Adey:2019ywk}, while $\chi^2_{\textrm{PRO}}$ is constructed similar to Eq.~(\ref{Eq:fitter}) by mapping the predicted ${}^{235}$U prompt energy spectrum $\boldsymbol{S}^{\rm fit}$ in Daya Bay to the predicted prompt energy spectrum in PROSPECT.
\iffalse

\begin{align} 
\label{Eq:fitter2}
\chi^2_{\textrm{PRO}} =& (\boldsymbol{R}^{\rm map}\boldsymbol{S}^{\rm fit}\eta^{\rm rate}-\boldsymbol{S}^{\rm PRO}_{\rm p})^{T}(\textbf{Cov}^{\rm PRO})^{-1} \nonumber\\
&(\boldsymbol{R}^{\rm map}\boldsymbol{S}^{\rm fit}\eta^{\rm rate}-\boldsymbol{S}^{\rm PRO}_{\rm p}).
\end{align} 
\fi
Importantly, inclusion of the unconstrained rate parameter $\eta^{\rm rate}$ introduces the shape-only constraint from PROSPECT into the Daya Bay deconvolution without biasing any absolute rate information.
For this shape-only analysis, the Daya Bay rate uncertainty is not included in uncertainties.
Daya Bay rate uncertainties are included in the latter part to extract the generic antineutrino energy spectra.

\begin{figure}
    \centering
    \includegraphics[width=\linewidth]{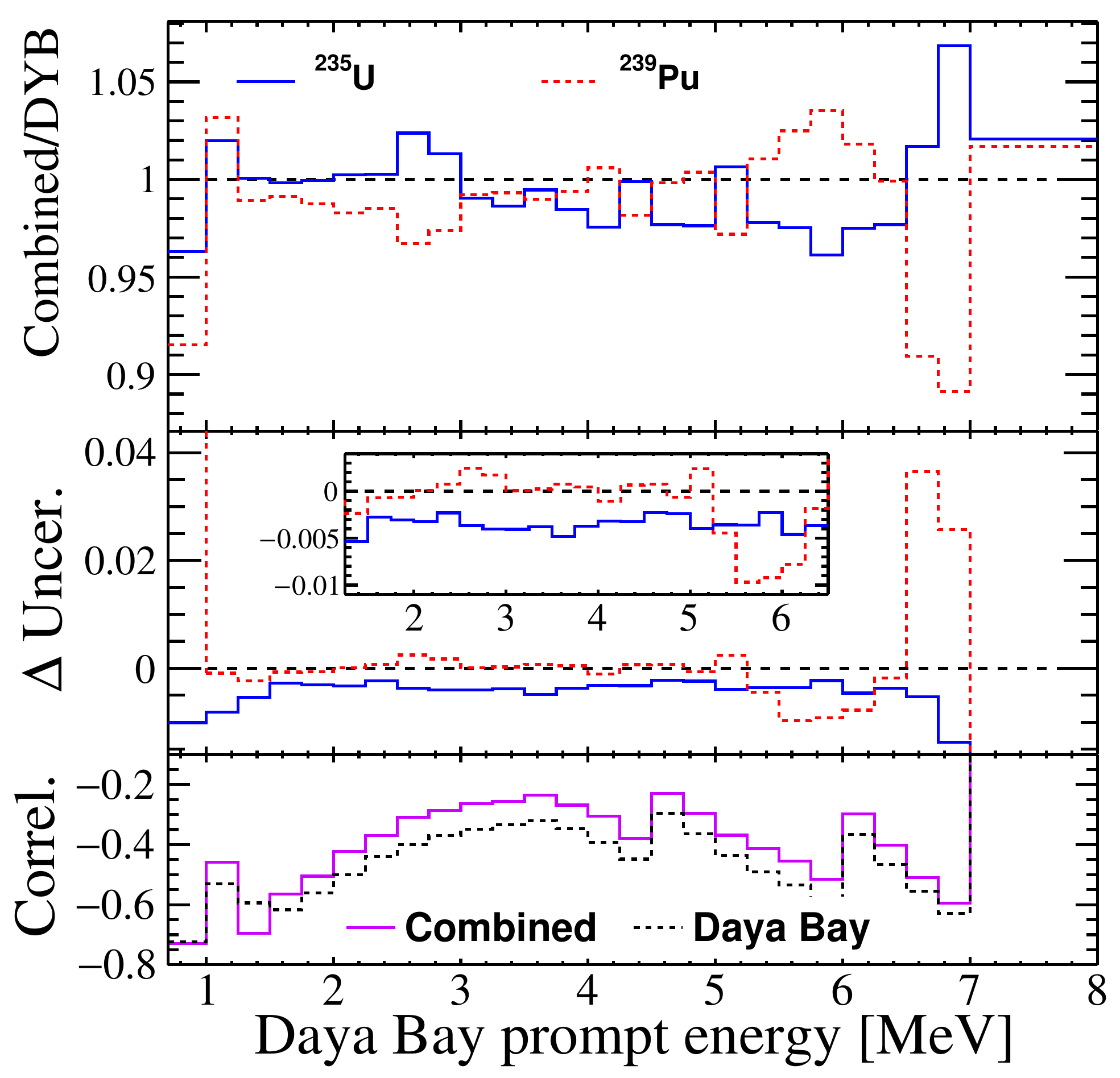}
    \caption{(Top) The ratio of the combined analysis results to the Daya Bay only results~\cite{Adey:2019ywk}. 
    (Middle) The difference of the relative uncertainties between the combined analysis results and the Daya Bay only results~\cite{Adey:2019ywk}. The inset shows the zoomed plot of the relative uncertainty differences. 
    (Bottom) The correlation coefficients of the extracted prompt energy spectra between ${}^{235}$U and ${}^{239}$Pu. 
    }
    \label{fig:JointAnalysisPromptCompare}
\end{figure}

The extracted ${}^{235}$U and ${}^{239}$Pu spectral shapes of the combined fit are shown in Fig.~\ref{fig:JointAnalysisPrompt}, 
and their difference from the previous result from Daya Bay~\cite{Adey:2019ywk} is shown in Fig.~\ref{fig:JointAnalysisPromptCompare}. 
The two results are consistent. 
With the additional constraints from PROSPECT data, the relative uncertainty of the spectral shape for ${}^{235}$U is improved from 3.5\% to 3\% around 3~MeV. 
The improvement in other energy regions is similar as shown in the middle panel of Fig.~\ref{fig:JointAnalysisPromptCompare}. 
The relative uncertainties of the spectral shape for ${}^{239}$Pu have no significant change.
However, the anticorrelation of the prompt energy spectra between ${}^{235}$U and ${}^{239}$Pu decreases by $\sim$20\% as shown in Fig.~\ref{fig:JointAnalysisPromptCompare}. 
With less degeneracy, the extracted ${}^{235}$U and ${}^{239}$Pu spectra change at the 2\% level compared with the results from Daya Bay alone, which is within the original 1$\sigma$ uncertainties. 

The extracted ${}^{235}$U and ${}^{239}$Pu spectral shapes are compared with the scaled Huber-Mueller model predictions as shown in Fig.~\ref{fig:JointAnalysisPrompt}. 
In the 4–6~MeV energy window, a 6\% (10\%) excess of events is observed for the $^{235}$U ($^{239}$Pu) spectrum compared with the prediction.
With Daya Bay data only, the local discrepancy between the extracted $^{235}$U ($^{239}$Pu) spectrum and its corresponding predicted spectrum in 2~MeV wide sliding energy windows is below 4.0$\sigma$ (1.2$\sigma$)~in Ref.~\cite{Adey:2019ywk}.  
With the combined measurement of Daya Bay and PROSPECT, the significance of the local deviation from the Huber-Mueller ${}^{235}$U model increases by $0.2\sigma$--$0.5\sigma$ at all energies, and the maximum local discrepancy increases to 4.2$\sigma$ around the 5~MeV prompt energy region. 
No significant change on the local deviation is observed for the ${}^{239}$Pu spectrum.

\begin{figure}
    \centering
    \includegraphics[width=\linewidth]{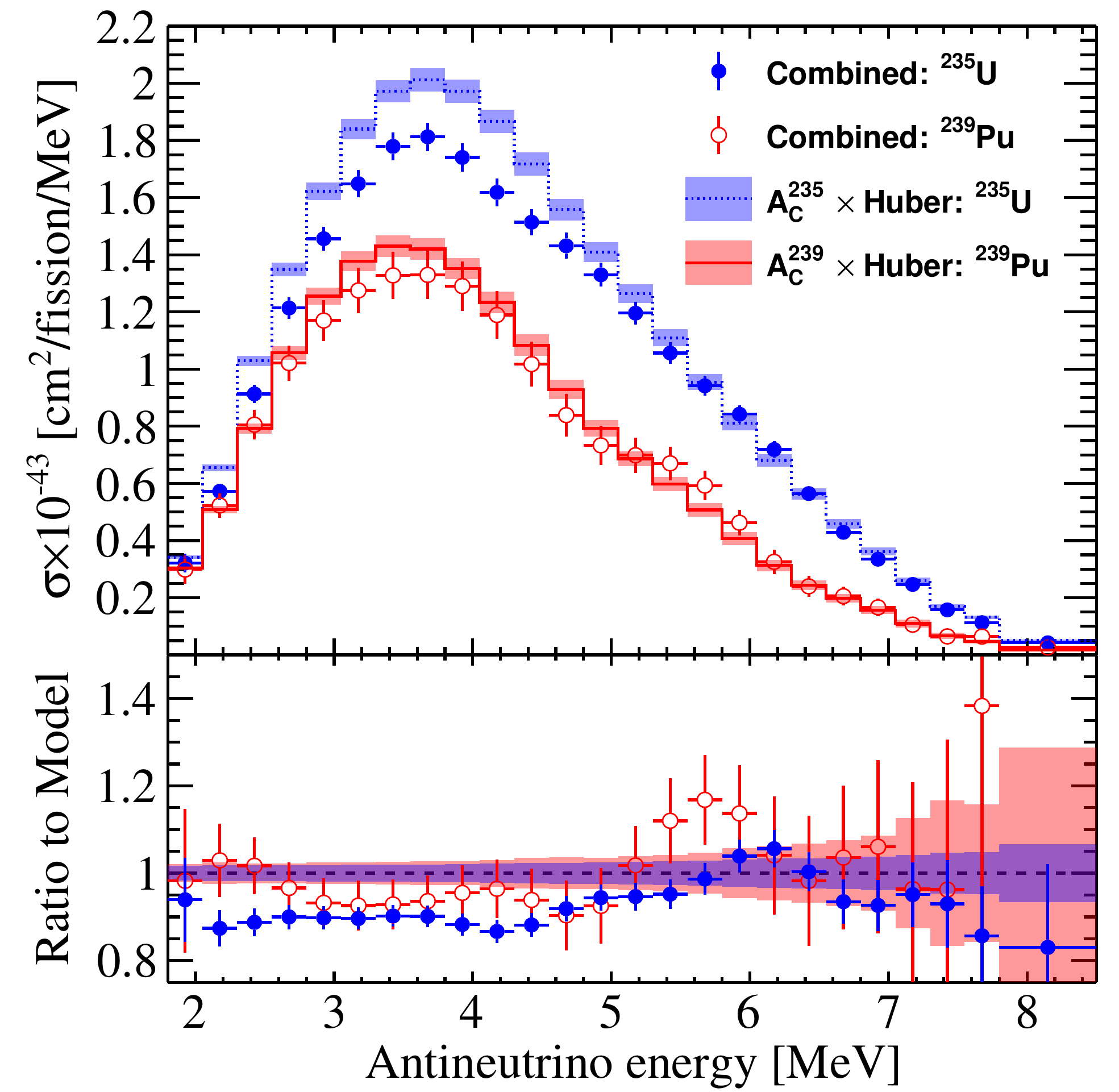}
    \caption{(Top) $^{235}$U and $^{239}$Pu antineutrino spectra unfolded from the jointly deconvolved Daya Bay and PROSPECT measurements.  (Bottom) Ratio of the measurements to their respective models, which are corrected by the smearing matrices $\boldsymbol{A}_c$ in both panels.}
    \label{fig:UnfoldedResults}
\end{figure}

Finally, $^{235}$U and $^{239}$Pu spectra expressed in antineutrino energy are obtained by unfolding the combined prompt energy spectra $\boldsymbol{S}^{\rm Com}_{\rm p}$ from the two experiments (shown in Fig.~\ref{fig:JointAnalysisPrompt}) using the Wiener-SVD unfolding technique~\cite{Tang:2017rob}, with analysis details similar to that in Ref.~\cite{DayaBay:2021dqj}.
For this portion of the analysis, the Daya Bay rate uncertainties are included.
Given the detector response matrix of Daya Bay $\boldsymbol{R}^{\rm DYB}$ and the covariance matrix $\textbf{Cov}^{\rm Com}$, 
the Wiener-SVD method derives:
\begin{equation} \label{eq:Unfold}
    \hat{\boldsymbol{S}}_{\bar{\nu}_e} = \boldsymbol{A}_C \cdot \big( \boldsymbol{\tilde{R}}^T\boldsymbol{\tilde{R}} \big)^{-1} \cdot \boldsymbol{\tilde{R}}^T \cdot \boldsymbol{Q} \cdot \boldsymbol{S}^{\rm Com}_{\rm p}, 
\end{equation}
where $\boldsymbol{\tilde{R}} = \boldsymbol{Q} \cdot \boldsymbol{R}^{\rm DYB}$ is the pre-normalized detector response matrix through the Cholesky decomposition $(\textbf{Cov}^{\rm Com})^{-1}=\boldsymbol{Q}^T \boldsymbol{Q}$. 
$\boldsymbol{A}_c$ is the smearing matrix obtained from the Wiener-SVD procedure to suppress noise fluctuations during unfolding process and maximize the signal-to-noise ratio in the effective frequency domain, allowing any model prediction to be smeared appropriately based on the regularization introduced by the unfolding.
The unfolded joint spectra are presented in Fig.~\ref{fig:UnfoldedResults} along with the Huber-Mueller prediction which has been smeared using $\boldsymbol{A}_c$.
The absolute rate deficit of data relative to the Huber-Mueller model is observed both in the full energy spectra and in the ratios in Fig.~\ref{fig:UnfoldedResults}.
The smearing matrices, unfolded spectra, and covariance matrices are included in the Supplemental Material. 
Examples demonstrating how to apply this smearing matrix and compare to a model are also given in the Supplemental Material and Ref.~\cite{DayaBay:2021dqj}.

In summary, the measured prompt IBD energy spectra of ${}^{235}$U by Daya Bay and PROSPECT are consistent.
A combined analysis between the two experiments is done and the results for ${}^{239}$Pu see no significant change, but uncertainties in the jointly determined spectral shape of the ${}^{235}$U prompt energy spectrum are reduced to 3\%. 
Additionally the degeneracy between ${}^{235}$U and ${}^{239}$Pu spectra is reduced by $\sim$20\%. 
This first combination of measurements from LEU and HEU reactors provides a more precise $\bar{\nu}_e$ energy spectrum for other reactor $\bar{\nu}_e$ measurements and other applications~\cite{Vivier:2019cot,Stewart:2019rtd,PhysRevApplied.9.014003,PhysRevApplied.8.034005}.  
The combined result can be further improved with increased statistics from Daya Bay, STEREO~\cite{AlmazanMolina:2020spe,AlmazanMolina:2020jlh}, the next generation of the PROSPECT experiment, and other complementary joint analyses~\cite{PhysRevLett.128.081802}.

%%%%%% Acknowledgements %%%%%%%%
The Daya Bay experiment is supported in part by the Ministry of Science and Technology of China, the U.S. Department of Energy, the Chinese Academy of Sciences, the CAS Center for Excellence in Particle Physics, the National Natural Science Foundation of China, the Guangdong provincial government, the Shenzhen municipal government, the China General Nuclear Power Group, the Research Grants Council of the Hong Kong Special Administrative Region of China, the Ministry of Education in Taiwan, the U.S. National Science Foundation, the Ministry of Education, Youth, and Sports of the Czech Republic, the Charles University Research Centre UNCE, the Joint Institute of Nuclear Research in Dubna, Russia, the National Commission of Scientific and Technological Research of Chile, We acknowledge Yellow River Engineering Consulting Co., Ltd., and China Railway 15th Bureau Group Co., Ltd., for building the underground laboratory. We are grateful for the ongoing cooperation from the China Guangdong Nuclear Power Group and China Light \& Power Company. 

The PROSPECT experiment is supported by the following sources: US Department of Energy (DOE) Office of Science, Office of High Energy Physics under Award No. DE-SC0016357 and DE-SC0017660 to Yale University, under Award No. DE-SC0017815 to Drexel University, under Award No. DE-SC0008347 to Illinois Institute of Technology, under Award No. DE-SC0016060 to Temple University, under Contract No. DE-SC0012704 to Brookhaven National Laboratory, and under Work Proposal Number  SCW1504 to Lawrence Livermore National Laboratory. 
This work was performed under the auspices of the U.S. Department of Energy by Lawrence Livermore National Laboratory under Contract DE-AC52-07NA27344 and by Oak Ridge National Laboratory under Contract DE-AC05-00OR22725. Additional funding for the experiment was provided by the Heising-Simons Foundation under Award No. \#2016-117 to Yale University. 

J.K.G. is supported through the NSF Graduate Research Fellowship Program and A.C. performed work under appointment to the Nuclear Nonproliferation International Safeguards Fellowship Program sponsored by the National Nuclear Security Administration’s Office of International Nuclear Safeguards (NA-241). 
This work was also supported by the Canada  First  Research  Excellence  Fund  (CFREF), and the Natural Sciences and Engineering Research Council of Canada (NSERC) Discovery  program under grant \#RGPIN-418579, and Province of Ontario.

We further acknowledge support from Yale University, the Illinois Institute of Technology, Temple University, Brookhaven National Laboratory, the Lawrence Livermore National Laboratory LDRD program, the National Institute of Standards and Technology, and Oak Ridge National Laboratory. We gratefully acknowledge the support and hospitality of the High Flux Isotope Reactor and Oak Ridge National Laboratory, managed by UT-Battelle for the U.S. Department of Energy.

\bibliography{references}% Produces the bibliography via BibTeX.

\end{document}